\documentclass[twocolumn,aps,prl,showpacs,amsmath,amssymb]{revtex4}
\usepackage{graphics}
\usepackage{epsf,graphicx,epsfig,psfrag}
\begin{document}

\title{Doping dependence of the Nernst effect in Eu(Fe$_{1-x}$Co$_x$)$_2$As$_2$ - departure from Dirac fermions physics}
\author{Marcin Matusiak$^1$, Zbigniew Bukowski$^2$, and Janusz Karpinski$^2$}
\affiliation{1.Institute of Low Temperature and Structural
Research, Polish Academy of Sciences, P.O. Box 1410, 50-950 Wroclaw, Poland}
\affiliation{2. Laboratory for Solid State Physics, ETH Zurich, 8093 Zurich, Switzerland} 
\date{\today}

\begin{abstract}
We report a systematic study of the transport properties in the series of  Eu(Fe$_{1-x}$Co$_x$)$_2$As$_2$ single crystals with $x$ = 0, 0.15, 0.20 and 0.30. Spin-density-wave order is observed in the undoped and the least doped samples ($x$ = 0, 0.15), while for $x$ = 0.15 and 0.20 Eu(Fe$_{1-x}$Co$_x$)$_2$As$_2$ becomes a superconductor. We found the properties of the parent EuFe$_2$As$_2$ compound well described by the Dirac fermions model, whereas cobalt doping caused an evolution of the system toward a regular metallic state. The antiferromagnetic ordering of the Eu$^{2+}$ ions at $T_N \approx$ 20 K has only minor influence on the measured quantities.
  
\end{abstract}


\maketitle

There is a substantial difference between the antiferromagnetic ground states of the parent compounds of copper- and iron-based superconductors. While the first is the Mott insulator \cite{Lee}, the spin-density-wave (SDW) state in the second is always metallic \cite{Sebastian}. Since the Cooper pairing interaction is probably magnetic in both families, understanding of the evolution of the  system from magnetism to superconductivity (SC) can be a crucial step towards revealing the mechanism responsible for superconductivity.
In this letter we investigate the Eu(Fe$_{1-x}$Co$_x$)$_2$As$_2$ series of iron-pnictide single crystals and report Nernst coefficient ($\nu$) data together with complementary studies of the thermoelectric power ($S$), Hall coefficient ($R_H$) and resistivity ($\rho$). The dominating influence of Dirac fermions on the transport properties seen in the parent EuFe$_2$As$_2$ compound vanishes with cobalt doping and our most highly doped Eu(Fe$_{0.7}$Co$_{0.3}$)$_2$As$_2$ shows regular metallic behavior. In the least doped Eu(Fe$_{0.85}$Co$_{0.15}$)$_2$As$_2$ we observe both superconductivity and spin-density-wave order. However, the influence of SDW on $\nu$ changes radically in comparison with the undoped EuFe$_2$As$_2$. This may indicate that Dirac fermions cannot survive in the sample that shows both SDW and SC.
  

Single crystals of Eu(Fe$_{1-x}$Co$_x$)$_2$As$_2$ were grown out of Sn flux. The constituent elements were loaded into alumina crucibles and placed in quartz ampoules sealed under pressure of 0.3 bar of Ar. The ampoules were heated to 1050$^o$C and kept at that temperature for 10 h to ensure complete dissolving of all components in molten Sn. Next, the ampoules were slowly (2-3$^o$C/h) cooled down to 600$^o$C, then liquid Sn-flux was decanted and remaining Sn was etched away from crystals with hydrochloric acid. To cover all possible SDW/SC configurations shown in Table 1, we selected four compositions for further studies: $x$=0 (denoted as Co-0), $x$=0.15 (Co-15), $x$=0.20 (Co-20), and $x$=0.30 (Co-30). The cobalt content was determined by the energy dispersive x-ray (EDX) analysis, which gave us values that were larger than the nominal and typical values from other studies \cite{Ying,Jiang,Nicklas}. We ascribe this to uncertainties arising from partial overlap of the main Eu, Fe and Co peaks in energy dispersive x-ray spectra. Therefore, the presented here absolute values of $x$ should be treated only as estimates.
 The phase purity was checked by powder X-ray diffraction (XRD). All the observed diffraction lines on the XRD pattern could be indexed on the basis of the tetragonal ThCr$_2$Si$_2$-type structure (space group I4/mmm). Both $a$ and $c$ lattice parameters show the systematic, but weak evolution with $x$ (see Table 1).
\begin{table}
\label{Tab1}
\caption{The lattice parameters and the influence of cobalt content on the presence/absence of the SDW and SC order in Eu(Fe$_{1-x}$Co$_x$)$_2$As$_2$.}
\begin{ruledtabular}
\begin{tabular}{c|c|c|c|c}
 $x$ & $a$ ($\AA$) & $c$ ($\AA$) & SDW, $T_{SDW}$ (K) & SC, $T_c$ (K) \\
\hline
 0 & 3.898(1)  & 12.11(1) & present, 191 K & absent, -\\
\hline
 0.15 & 3.904(1)  & 12.08(1) & present, 131 K & present, 7.7 K\\
\hline
 0.20 & 3.911(1)  & 12.06(1) & absent, - & present, 5.2 K\\
\hline
 0.30 & 3.912(1)  & 12.03(1) & absent, - & absent, -\\
\end{tabular}
\end{ruledtabular}
\end{table}

The methods of measurements of the electrical resistivity, Hall coefficient, thermoelectric power, and Nernst coefficient were the same as described in Ref. \cite{Matusiak} with one important difference. In Ref. \cite{Matusiak} we used the old sign convention, according to which the vortex Nernst signal is negative, whereas in the present paper we use a recently more popular convention \cite{Behnia}. This results in opposite Nernst coefficients of CaFe$_{2-x}$Co$_x$As$_2$ \cite{Matusiak} and Eu(Fe$_{1-x}$Co$_x$)$_2$As$_2$, despite the fact that Nernst signals in both materials have the same sign.


The temperature dependences of the electrical resistivity shown in Fig.1 reveal the emergence of the SDW state, which is accompanied by the structural transition \cite{Tegel}, at $T_{SDW}$ = 191 K for Co-0 and $T_{SDW}$ = 131 K for Co-15.
\begin{figure}
\label{Fig1}
 \epsfxsize=8.5cm \epsfbox{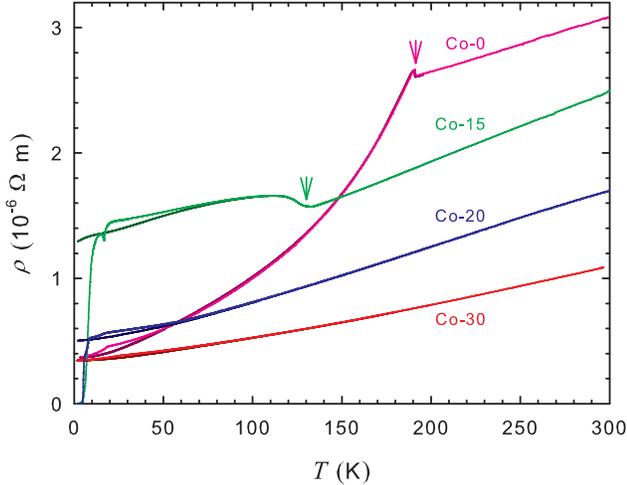}
 \caption{(Color online) The temperature dependences of the resistivities of the Eu(Fe$_{1-x}$Co$_x$)$_2$As$_2$ series. At low temperatures the lines deviating upward are measured at $B=0$ T, while their featureless counterparts are measured in field of 12.5 or 13 T. Arrows indicate the onset of the SDW order in Co-0 and Co-15.}
 \end{figure}
Additionally, we observe the superconducting transition in Co-15 and Co-20 at $T_c$ = 7.7 K and $T_c$ = 5.2 K, respectively ($T_c$ is defined as the maximum in $d\rho/dT$). At low magnetic field ($B \lesssim $ 0.5 T) the onset of the superconducting transition in the Co-15 crystal is notably above $T_c$, but below 17 K the resistivity temporarily goes back to its normal value as superconductivity is destroyed by the competing antiferromagnetic order of the Eu$^{2+}$ ions \cite{Jeevan1}. Analogous reentrant behavior was already reported for Eu(Fe$_{1-x}$Co$_x$)$_2$As$_2$ \cite{Jiang,Nicklas}, EuFe$_2$(As$_{1-x}$P$_x$)$_2$ \cite{Ren,Jeevan2}, and the undoped EuFe$_2$As$_2$ under pressure \cite{Miclea}. The Eu$^{2+}$ ordering is visible in the all studied samples as a small and broad peak in $\rho(T)$ around $T$ = 20 K. This peak is completely eradicated by magnetic field of the order of 10 T. This happens irrespectively of the $B$ vector orientation (for Co-20 $B$ is parallel to the {\bf c} crystallographic axis, for all other samples $B$ is perpendicular to {\bf c}). Such a magnetic field is also sufficient to completely suppress the superconducting transition, or at least, to shift $T_c$ below $T\approx$ 2 K.
Fig. 2 presents the temperature and doping dependences of the Hall coefficient ($B\approx$ 13 T) - panel ($a$), thermoelectric power ($B$ = 0 T) - panel ($b$), and the Nernst coefficient ($B\approx$ 13 T) - panel ($c$) for the Eu(Fe$_{1-x}$Co$_x$)$_2$As$_2$ series.
\begin{figure}
\label{Fig2}
 \epsfxsize=9cm \epsfbox{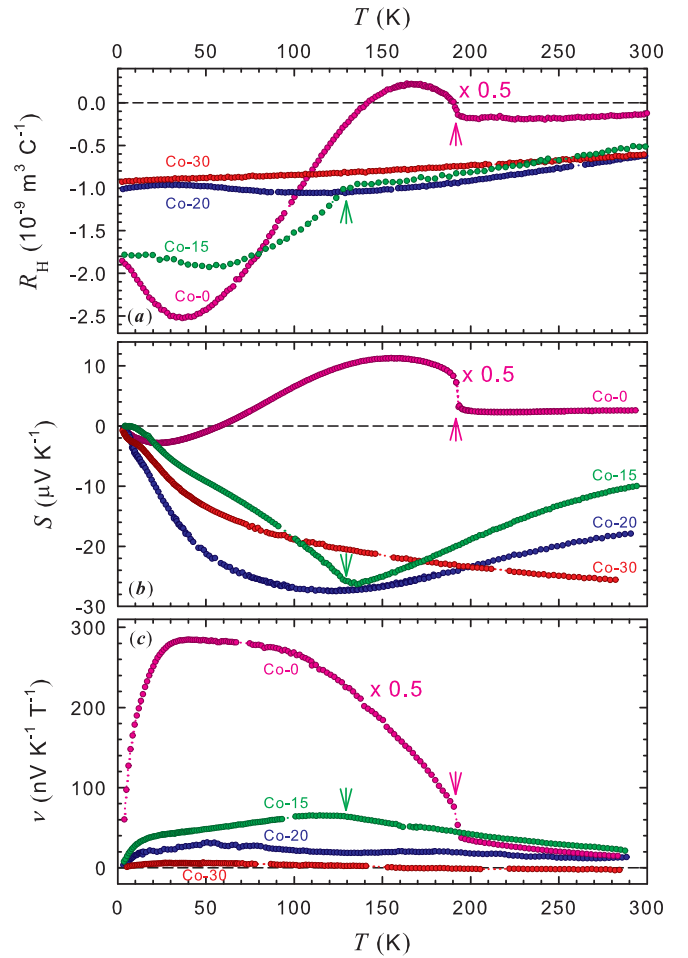}
 \caption{(Color online) The temperature dependences of the Hall coefficient (panel $a$), thermoelectric power (panel $b$) and Nernst coefficient (panel $c$)  for the Eu(Fe$_{1-x}$Co$_x$)$_2$As$_2$ series. All coefficients for Co-0 are divided by 2. Arrows indicate $T_{SDW}$ for Co-0 and Co-15.}
 \end{figure}
The high temperature properties of all measured quantities systematically evolve with increasing $x$ towards the characteristics of a regular metal represented by Co-30, the crystal with the highest doping. For this sample $R_H$ is small and weakly temperature dependent, $S$ is nearly linear with $T$, and $\nu$ becomes very small ( $|\nu| < 5$ nV K$^{-1}$ T$^{-1}$) as expected in the case of the satisfied Sondheimer cancellation \cite{Sondheimer}. The Nernst coefficient at zero temperature can be related to the Fermi temperature ($T_F$) and Hall mobility ($\mu_H \equiv \frac{\sigma_{xy}}{B\sigma_{xx}} = \frac{R_H}{\rho}$) through the equation: $\nu = \frac{\pi2 k_B}{3e} \frac{T \mu_H}{T_F}$ \cite{Behnia}, where $k_B$ is the Boltzmann constant and $e$ is the elementary charge. As seen in Fig. 3, low temperature values of $\nu/T$ saturate for all samples and can be used to estimate the Fermi energy. The approximative (Eu(Fe$_{1-x}$Co$_x$)$_2$As$_2$ is a multi-band system) values of $T_F$ together with $\nu/T$ and $\mu_H$ are collected in Tab. 2.
\begin{figure}
\label{Fig3}
 \epsfxsize=8.5cm \epsfbox{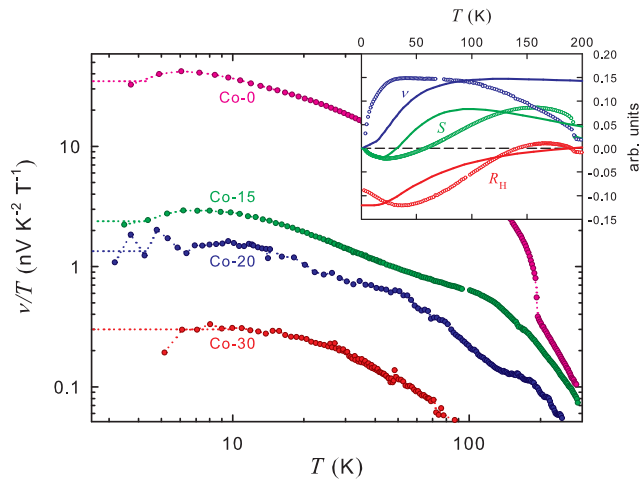}
 \caption{(Color online) The magnitude of the Nernst coefficient divided by
temperature for the Eu(Fe$_{1-x}$Co$_x$)$_2$As$_2$ series plotted versus temperature on a logarithmic scale. Dotted lines are guides for the eye. Inset shows temperature dependences of transport coefficients for Co-0 compared with theoretical results from Ref. \cite{Morinari}.}
 \end{figure}
 
\begin{table}
\label{Tab2}
\caption{Summary of results at the low temperature limit.}
\begin{ruledtabular}
\begin{tabular}{c|c|c|c}
 $x$ & $\mu_H$ (T$^{-1}$) & $\nu / T$ (nV K$^{-2}$ T$^{-1}$) & $T_F$ (K) \\
\hline
 0 & -0.0098 & 34 & 80\\
\hline
 0.15 & -0.0014 & 2.3 & 170\\
\hline
 0.20 & -0.002 & 1.3 & 440\\
\hline
 0.30 & -0.0027 & 0.3 & 2500\\
\end{tabular}
\end{ruledtabular}
\end{table}
Intriguingly, the crystal with the lowest Fermi energy ($\varepsilon_F \approx$ 7 meV) has at low temperatures clearly the highest $\mu_H$, and such a significantly enhanced mobility can be a manifestation of limited scattering of the Dirac fermions. Their presence in the SDW phase of the iron-pnictides was theoretically predicted \cite{Ran}, and suggested to play an important role in transport properties \cite{Morinari,Huynh}. Moreover, a Dirac cone was observed in the electronic structure of BaFe$_2$As$_2$ by angle resolved photoemission spectroscopy (ARPES) \cite{Richard} and was shown to be consistent with the angle dependence of the magnetic quantum oscillations in BaFe$_2$As$_2$ and SrFe$_2$As$_2$ \cite{Harrison}. Recent theoretical investigations of Dirac fermions in the parent antiferromagnetic state by Morinari et al. \cite{Morinari} have provided $R_H(T)$, $S(T)$ and $\nu(T)$, whose overall trends agree well with the experimental data presented here (the authors of Ref. \cite{Morinari} used the same "old" sign convention as in Ref. \cite{Matusiak}). The authors considered a phenomenological two-band model consisting of a hole band (denoted below with index $h$) with a conventional energy spectrum, and an electron band (index $e$) with the Dirac fermion energy spectrum. It is worth emphasizing that even if the Dirac fermions are the minority carriers some transport coefficients exhibit noticeable contributions from Dirac fermions and the Nernst coefficient is expected to be significantly affected \cite{Morinari}. $\nu$ was calculated as: $\nu = (\alpha_{xx}\sigma_{xy}-\alpha_{xy}\sigma_{xx})/[B (\sigma_{xx}^2 + \sigma_{xy}^2)]$, the Hall coefficient as: $R_H = \sigma_{xy}/(B \sigma_{xx}^2)$ and the thermopower as: $S = \tau \alpha_{xx}$, where $\alpha_{ij}$ and $\sigma_{ij}$ are elements of the Peltier and electrical conductivity tensors, respectively, and $\tau$ is the relaxation time. Results are the sum of contributions from the two bands. The inset to Fig. 3 shows the comparison between the theoretical and the experimental data (Co-1), where the latter were multiplied by a constant to match the heights of maxima or minima. The theoretical curves are an exact copy of the results presented by Morinari et al. (Fig. 3($c$)) \cite{Morinari} obtained for relaxation times ratio $\tau_h/\tau_e= 0.45$, concentrations ratio $n_e/n_h = 0.05$, and the value of $\varepsilon_0=\varepsilon_e$= $k_B T_F$ was taken from Table 2 above. 
What we would like to stress here is that the response of the electronic system to the onset of SDW is different in Co-0 and Co-15. Furthermore, for the Nernst coefficient this difference is substantial. Namely, in Co-0 there is a sudden rise of $\nu$ below $T_{SDW}$ that is very similar to one reported in CaFe$_2$As$_2$ \cite{Matusiak}, while in Co-15 $\nu$ for $T<T_{SDW}$ decreases slightly below the high temperature $\nu(T)$ trend. Fig. 4 shows an attempt to separate this anomalous and normal contributions to the Nernst signal.
\begin{figure}
\label{Fig4}
 \epsfxsize=8.5cm \epsfbox{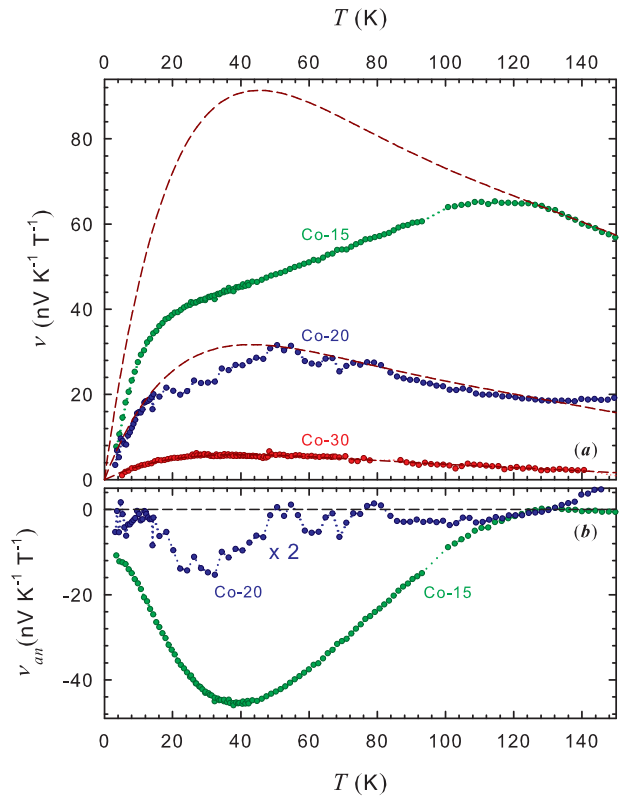}
 \caption{(Color online) Separation of $\nu(T)$ in Co-15 and Co-20 into the normal $(\nu_{met})$ and anomalous $(\nu_{an})$ components. Panel $(a)$ shows the experimental data (points) and fitted $\nu_{met}(T)$ (dashed lines). Panel $(b)$ presents the estimated $\nu_{an}(T)$ for Co-15 and Co-20 (the latter is multiplied by 2).}
 \end{figure}
To this end we utilized the purely metallic $\nu_{met}(T)$ dependence of Co-30, which was fitted to the high temperature part of $\nu(T)$ of Co-15 and also Co-20. Fittings were made with two free parameters: $\nu_{met}(T) = aT + b \nu_{Co-30}(T)$, where $a$ and $b$ were supposed to provide for variation of the of the Nernst coefficients due to modification of scattering and concentration of the charge carriers by Co-doping. The total Nernst coefficient is assumed to simply be a sum of the normal $(\nu_{met})$ and anomalous $(\nu_{an})$ parts: $\nu(T)=\nu_{met}(T)+\nu_{an}(T)$. $\nu_{an}(T)$ obtained in this way for Co-15 and Co-20 are presented in Fig. 4 ($b$). The onset of the anomaly in Co-15 correlates with $T_{SDW}$, and is likely associated with the Fermi surface reconstruction caused by spin modulations. A description of the normal-state Nernst signal in the presence of spin-only, charge-only, and combined spin and charge stripe order was recently proposed by Hackl et al. \cite{Hackl}. The authors employed a phenomenological quasiparticle model combined with a Boltzmann equation approach and showed that Fermi pockets caused by translational symmetry breaking can lead to an enhancement of $\nu$. The sign of the anomaly depends on the strength as well as the period of the modulation. Since stripe fluctuations were suggested to be sufficient to cause the Nernst coefficient to increase \cite{Cyr,Taillefer}, a small anomaly in $\nu(T)$ that is present in the Co-20 crystal below $T$ = 50 K might be attributed to SDW fluctuations. However, a very limited size of this maximum does not allow us to draw definitive conclusions.

While the onset of the SDW order has undoubtedly a significant impact on the electronic transport properties, the local magnetic order in the Eu sublattice seems to have no effect on the Nernst signal. Figure 5 shows the $\nu(T)$ dependences for Co-15 measured at various $B$.
\begin{figure}
\label{Fig5}
 \epsfxsize=8.5cm \epsfbox{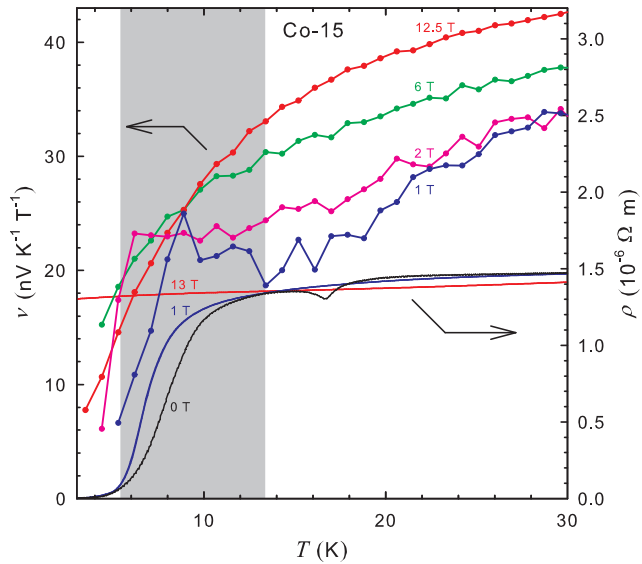}
 \caption{(Color online) The temperature dependences of the Nernst coefficient (left axis) and resistivity (right axis) for Co-15 measured at various magnetic fields. The shaded area denotes the temperature range, where $\nu(T)$ at $B$ = 1 T seems to be influenced by superconducting fluctuations.}
 \end{figure}
A positive contribution to $\nu$ visible at $B$ = 1 T and 5 K $< T <$ 13 K correlates with the superconducting transition and most probably is a trace of vortex movement. There is no other anomaly at low fields ($B = 1$ and 2 T), where the influence of the antiferromagnetic Eu$^{+2}$ ordering on the resistivity is still noticeable. A field dependent $\nu$ seems to be related to the presence of the SDW order and it is also observed in the non-superconducting Co-0 crystal. We were unable to detect the nonlinearity of $\nu(B)$ in Co-20 and there is no detectable Nernst signal from the SC fluctuations/vortices as in the previously reported case of Ca(Fe$_{0.96}$Co$_{0.04}$)$_2$As$_2$ \cite{Matusiak}. In Co-30 the nonlinearities in $\nu(B)$ are absent as well. These results confirm the weak electronic coupling between the Eu and FeAs sublattices \cite{Zhou,Terashima}.

In summary, we have presented data indicating that the low temperature transport properties of the EuFe$_2$As$_2$ parent compound are dominated by Dirac fermions thus this compound can be considered as a nodal SDW material. Co doping causes the sudden change of characteristic in the SDW state and the influence of the Dirac fermion vanishes in the superconducting Eu(Fe$_{0.85}$Co$_{0.15}$)$_2$As$_2$. An open question is whether this is a consequence of changes of tiny electron pockets induced by a shift of the Fermi level, or rather increased scattering Dirac fermions, or perhaps an interaction between the nodeless s$^{\pm}$ type superconductivity and gapless Dirac fermions \cite{Hasan}. 

\section*{Acknowledgments}
The authors are grateful to L.J. Spalek, A. Hackl, J.R. Cooper and T. Morinari for useful comments and to M. Malecka for performing the EDX analysis. This work was supported by a grant No. N N202 130739 of the Polish Ministry of Science and Higher Education.

\end{document}